\documentclass[a4paper]{jpconf}
\usepackage{graphicx}
\usepackage{physics}
\usepackage{tensor}
\usepackage{float}
\usepackage[position=top]{subfig}
\usepackage[labelfont=bf]{caption}
\usepackage{bm}
\usepackage{siunitx}
\usepackage{multicol}
\usepackage{soul}
\usepackage{pgfgantt}
\usepackage{mathtools}
\usepackage{amsmath}
\usepackage{cancel}
\usepackage{enumitem}
\usepackage{etoolbox}
\usepackage{booktabs,arydshln,ltablex}
\usepackage{adjustbox}
\usepackage{multirow}

\usepackage{tikz}

\begin{document}

\title{Relativistic Anisotropic Polytropic Spheres: Physical Acceptability}

\author{Daniel Suárez-Urango$^{1}$, Luis A. Núñez$^{1,2}$ and Héctor Hernández$^{1,2}$}

\address{$^1$ Escuela de Física, Universidad Industrial de Santander, Bucaramanga, Colombia}
\address{$^2$ Departamento de Física, Universidad de los Andes, Mérida, Venezuela}

\ead{danielfsu@hotmail.com}

\begin{abstract}
In this work we evaluate the physical acceptability of relativistic anisotropic spheres modeled by two polytropic equations of state -with the same newtonian limit- commonly used to describe compact objects in General Relativity. We integrate numerically the corresponding Lane-Emden equation in order to get density, mass and pressure profiles. An ansatz is used for the anisotropic pressure allowing us to have material configurations slightly deviated from isotropic condition. Numerical models are classified in a parameter space according to the number of physical acceptability conditions that they fulfil. We found that the polytropes considering total energy density are more stable than the second type of polytropic EoS.
\end{abstract}

\section{Introduction}
The analysis of the presence and propagation of instabilities in compact objects has been the subject of research for decades.  Only those stable configurations can represent real entities of astrophysical interest. Typically, stars are modelled as spherical objects --with gravity as the only binding force-- using structure equations that guarantee their hydrostatic equilibrium. These astrophysical objects' properties are described by an equation of state (EoS) relating their thermodynamic physical variables.

The polytropic EoS $P = K \rho^{1 + 1/n}$, relates pressure and density through a power law and has been fundamental as an approximation to realistic EoS \cite{Chandrasekhar1967,Tooper1964,HerreraBarreto2013}. It is fascinating since it can model a wide variety of different astrophysical scenarios by only varying the polytropic index $n$. From the General Relativity framework, there are two classes of polytropic EoS that converge in the Newtonian limit \cite{HerreraBarreto2013}: the first EoS relates pressure to rest mass density while the second relates the pressure to the total energy density.

This work determines the critical parameter that condition the stability of anisotropic hydrostatic material configurations. Anisotropy concerns non-Pascalian fluids, i.e. those having unequal radial and tangential pressures distributions within the matter configuration. Thus, $\Delta = P - P_{\perp}$ is the factor that measures the deviation from the isotropy condition $(P = P _ {\perp})$ with spherical symmetry -modelled by polytropic EoS- through the compliance of 9 physical acceptability conditions. These conditions range from restrictions on the energy-momentum tensor (energy conditions),  condition on the metric potentials at the boundary and sub luminous speeds of sound, passing through the stability criterion for an adiabatic index, up to more recent criteria such as cracking against local density perturbations and convective stability \cite{HernandezNunezVasquez2018}.

The density profiles are obtained by integrating the relativistic Lane-Emden equation numerically. As a final result, we present a parameter space generated by a set of stable models that fulfil all the acceptability conditions.

This work is organized as follows. The next section displays the notations and the basic assumption for the metric, the energy-momentum tensor, and the relativistic stellar structure equations. Section \ref{AcceptabilityConditions} discusses the acceptability conditions for relativistic compact objects, and section \ref{PolytropicEoS} reviews the theory of both relativistic polytropic EoS. We present the two systems of Lane-Emden structure equations in Section \ref{LaneEmden}. Next, in Section \ref{Result} we numerically integrate both systems and evaluate each of the acceptability conditions for different stiffness values, the anisotropic parameter and the polytropic index. Finally, Section \ref{Conclusions} ends with some general remarks.  

\section{Modeling Relativistic Anisotropic Spheres}
Let us consider spherical, static, anisotropic distribution of matter described by the line element
\begin{equation}
\mathrm{d}s^2 = {\rm e}^{2\nu(r)}\,\mathrm{d}t^2-{\rm e}^{2\lambda(r)}\,\mathrm{d}r^2- r^2 \left(\mathrm{d}\theta^2+\sin^2(\theta)\mathrm{d}\phi^2\right)\, .
\label{metricSpherical}
\end{equation}

The momentum-energy tensor for a non-Pascalian fluid with energy density $\rho$, radial pressure $P$ and tangential pressure $P_{\perp}$ is
\begin{equation}
T_\mu^\nu = \mbox{diag}\left[\rho(r),-P(r),-P_\perp(r),-P_\perp(r)  \right] \,,
\label{tmunu}
\end{equation}
and the corresponding Einstein Field Equations can be written as
\begin{eqnarray}
\rho(r)&=& \frac{ {\rm e}^{-2\lambda}\left(2 r \lambda^{\prime}-1\right)+1 }{8\pi r^{2}}\,,\label{FErho} \\
P(r) &=&  \frac {{{\rm e}^{-2\,\lambda}}\left(2r \,\nu^{\prime} +1\right) -1}{8 \pi\,{r}^{2}}\,\label{FEPrad} \qquad \textrm{and} \\
P_\perp(r) &=&-\frac{{\rm e}^{-2\lambda}}{8\pi}\left[ \frac{\lambda^{\prime}-\nu^{\prime}}r-\nu^{\prime \prime }+\nu^{\prime}\lambda^{\prime}-\left(\nu^{\prime}\right)^2\right] \label{FEPtan} \,, 
\end{eqnarray}
where prime denotes derivative with respect to $r$.

Now, redifining the metric function $\lambda(r)$ in terms of the Misner ``mass function'' we get \cite{MisnerSharp1964}
\begin{equation}
 m(t,r)=\frac{r^2}{2}R^{3}_{232} \; \Leftrightarrow \; m(r,t)=4\pi \int ^r_0 T^0_0r^2\mathrm{d}r \; \Rightarrow e^{-2\lambda}= 1-\frac{2 m(r,t)}{r}, 
\label{MassDef} 
 \end{equation}
and from (\ref{FEPrad}) we can write
\begin{equation}
\label{nu}
\nu^{\prime} = \frac{m + 4 \pi r^{3} P}{r(r - 2m)}.
\end{equation}

Finally, the hydrostatic equilibrium equation --the Tolman-Oppenheimer-Volkoff equation, i.e. $T^{\mu}_{r \; ; \mu}~=~0$-- for this anisotropic fluid can be obtained by differentiating equation (\ref{FEPrad}) with respect to $r$ and replacing the result in (\ref{FEPtan}), yielding
\begin{equation}
\frac{\mathrm{d} P}{\mathrm{d} r} = -(\rho +P)\frac{m + 4 \pi r^{3}P}{r(r-2m)} +\frac{2}{r}\left(P_\perp-P \right) \,.
 \label{TOVStructure1}
\end{equation}
Assuming a particular form for the  anisotropy pressure \cite{CosenzaEtal1981}, as
\begin{equation}
   \Delta\equiv P_{\perp} - P = C r (\rho + P)\left[ \frac{m + 4 \pi r^3 P}{r(r-2m)}\right]\,, 
   \label{DelCosenza}
\end{equation}
 we can write (\ref{TOVStructure1}) as
\begin{equation}
    \frac{\mathrm{d}P}{\mathrm{d}r} = - h \frac{(\rho + P)(m + 4 \pi  r^3 P)}{r(r-2m)}\,,
\label{ansatzcosenza}
\end{equation}
where $h = 1 - 2C$, with $C$ quantifying the deviation from the isotropic condition $(C = 0)$. 

\section{Physical Acceptability Conditions}
\label{AcceptabilityConditions}
Acceptability conditions are crucial concepts when considering self-gravitating stellar models. Only acceptable objects are of astrophysical interest, for this they must to comply with a set of acceptability conditions which can be stated  as \cite{ HernandezNunezVasquez2018, Ivanov2017, HernandezNunezSuarez2020}:
\begin{enumerate}
\item[{\bf C1}] $2m/r < 1$.

\item[{\bf C2}] Positive density and pressures, finite at the center of the configuration with $P_c=P_{\perp c}$.

\item[{\bf C3}] $\rho^{\prime} < 0$, $P^{\prime} < 0$, $P_{\perp}^{\prime} < 0$ with density and pressures having maximums at the center, thus $\rho^{\prime}_{c}=P^{\prime}_{c} = P^{\prime}_{\perp c}=0$,  with $P_{\perp} \geq P$.

\item[{\bf C4}] The strong energy condition for imperfect fluids, $\rho - P - 2P_{\perp} \geq 0$.

\item[{\bf C5}]  The dynamic perturbation analysis restricts the adiabatic index 
\[
\Gamma = \frac{\rho + P}{P} v_s^{2} \geq \frac{4}{3} \,.
\]

\item[{\bf C6}] Causality conditions on sound speeds: $0 < v_{s}^2 \leq 1$ and $0 < v_{s \perp}^2 \leq 1$.

\item[{\bf C7}] The Harrison-Zeldovich-Novikov stability condition: $\mathrm{d}M(\rho_c)/\mathrm{d}\rho_c > 0$.

\item[{\bf C8}] Cracking instability against local density perturbations, $\delta \rho = \delta \rho(r)$.  

\item[{\bf C9}] The adiabatic convective stability condition  $\rho^{\prime \prime} \leq 0$.
\end{enumerate}

\section{Polytropic Equations of State}
\label{PolytropicEoS}
Polytropic EoS have been widely used in newtonian physics to model several different scenarios \cite{goldreich1980,abramowicz1983polytropes,kovetz1968slowly,horedt2004polytropes}. It can be expressed as
\begin{equation}
\label{Poli1}
    P = K \hat{\rho}^{\gamma} = K \hat{\rho}^{1+1/n} \, ,
\end{equation}
where $\hat{\rho}$ is the baryonic mass density, while  $K$, $\gamma$, and  $n$, are the polytropic constant, polytropic exponent and polytropic index, respectively.

On the other hand, in General Relativity, there appear two possibilities for polytropic EoS depending on the type of density we are considering, i.e., energy density or baryonic mass density. These two equations lead to the same Newtonian limit  (\ref{Poli1}) \cite{HerreraBarreto2013}.

\subsection{Polytropic EoS I}
We can write the first and second law of thermodynamics as
\begin{equation*}
    \mathrm{d}\left(\frac{\rho + P}{\mathcal{N}} \right) -\frac{\mathrm{d}P}{\mathcal{N}} = T\mathrm{d}\left(\frac{\mathcal{S}}{\mathcal{N}} \right) \, \, ,
\end{equation*}
where $T$ is temperature, $\mathcal{S}$ is entropy per unit of proper volume and $\mathcal{N}$ is the particle density such that $\hat{\rho} = \mathcal{N}m_0$. Then, for an adiabatic process we have
\begin{equation}
\label{primeraley}
    \mathrm{d}\left(\frac{\rho + P}{\mathcal{N}} \right) -\frac{\mathrm{d}P}{\mathcal{N}} = 0 \, \, \Rightarrow \, \,  \mathrm{d} \left( \frac{\rho}{\mathcal{N}} \right) + P\mathrm{d} \left(\frac{1}{\mathcal{N}}  \right) = 0 \, \,.
\end{equation}
Now, using polytropic EoS (\ref{Poli1}) and knowing that $\mathcal{N} = \hat{\rho}/m_0$, yields
\begin{equation*}
    K \hat{\rho}^{\gamma - 2}  = \frac{\mathrm{d}\left(\rho /\hat{\rho} \right)}{\mathrm{d}\hat{\rho}}.
\end{equation*}
Considering $\gamma \neq 1$ we can integrate the former equation resulting
\begin{equation*}
    \rho = C_{1} \hat{\rho} + \frac{P}{\gamma - 1} \, ,
\end{equation*}
where $C_1$ is a constant equal to 1 since in the non-relativistic limit $\rho \rightarrow \hat{\rho}$. Thus
\begin{equation}
    \rho = \hat{\rho} + \frac{P}{\gamma - 1} = \hat{\rho} + n P \, .
\label{RelacionDensidades}
\end{equation}

\subsection{Polytropic EoS II}

Now, supposing the polytropic relation is
\begin{equation}
\label{Poli2}
    P = K \rho^{1+1/n},
\end{equation}
and using (\ref{primeraley}) and (\ref{Poli2}) we have
\begin{equation}
\frac{\mathrm{d} \rho}{\mathrm{d}{\hat \rho}} = \frac{K{\rho}^{\gamma} + \rho}{\hat \rho}  \,,
\end{equation}
yielding
\begin{equation}
    \int \frac{\mathrm{d} \rho}{K{ \rho}^{\gamma} + \rho}=\ln{\left(\frac{{\hat \rho}}{C_{2}}\right)} \qquad \textrm{for} \quad \gamma \neq 1 \,.
\end{equation}
Solving the former equation we get
\begin{equation*}
    \rho = \frac{\hat{\rho}}{\left(1 - K \hat{\rho}^{1/n}\right)^{n}} \, ,
\end{equation*}
where $C_2$, again, is a constat equal to 1.

\section{Lane-Emden Equation}
\label{LaneEmden}
The Lane-Emden equation is a dimensionless form of the hydrostatic equilibrium equation for polytropic fluids. 

\subsection{Lane-Emden equation for polytropic EoS I}
Changing to dimensionless variables
\begin{equation}
\label{CVmasayradio}
     \eta \left(\xi \right) = \frac{m}{4 \pi \rho_c a^{3}} \quad \textrm{and} \quad r = a\xi \qquad \textrm{where}
     \quad a^{2} = \frac{\sigma \left(n + 1 \right)}{4 \pi \rho_c} \quad \textrm{and} \quad     \hat{\psi}^{n} \left(\xi \right) = \frac{\hat{\rho}}{\hat{\rho}_{c}}.
\end{equation}
Thus the structure equations for a relativistic polytropic anisotropic matter distribution, now become 
\begin{eqnarray}
    \dot{\hat{\psi}} &=& \frac{h \left(\eta + \sigma \xi^{3} \hat{\psi}^{n+1} \right) \left[ 1 - \sigma n  + \sigma \left(n+1 \right) \hat{\psi} \right]}{\xi \left[2 \sigma \left(n+1 \right) \eta - \xi \right]} \label{LEIa}  \qquad \textrm{and} \\
    \dot{\eta} &=& \xi^{2} \hat{\psi}^{n} \left(1 - \sigma n  + \sigma n \hat{\psi} \right) \label{LEIb} \,,
\end{eqnarray}
where $\sigma = P_{c}/\rho_{c}$ and the dots indicate derivative with respect to $\xi$.

\subsection{Lane-Emden equation for polytropic EoS II}
With the same variables (\ref{CVmasayradio}) but using 
$    \psi^{n}(\xi) = \frac{\rho}{\rho_{c}} $, 
we have for energy density:
\begin{eqnarray}
    \dot{\psi} &=& \frac{h(\eta + \sigma \xi^{3} \psi^{n+1})(1 + \sigma \psi)}{\xi \left[2 \sigma (n+1) \eta - \xi \right]} \qquad \textrm{and} \label{LEIIa} \\
    \dot{\eta} &=& \xi^{2}\psi^{n} \label{LEIIb} \,.
\end{eqnarray}

 Numerical integration of equations (\ref{LEIa}), (\ref{LEIb}) and  (\ref{LEIIa}), (\ref{LEIIb})  was performed in $Python$, using the $RK45$ method with $solve\_ivp$ routine, with an accuracy of $10^{-15}$ for vanishing pressure at the boundary surface. The initial conditions of the two systems of equations given by $\hat{\psi} (\xi = 0) = \psi (\xi = 0) = 1 \,,  \, \eta (\xi = 0) = 0  \,.$
 
 Each solution can be labelled with a triplet of parameters, namely, polytropic index $n$, the ratio between pressure and density at centre $\sigma$, and anisotropic factor $C$. The variation of these variables gives a wide range of parameter space models, depending on the number of physical acceptability conditions they fulfil. 
 
 Polytropic index $n$ allows to model different kind of matter within the sphere. It is well known that $n = 0$ is associated with an incompressible fluid \cite{Bludman1973}, while $n = 3$ is used to model an utterly degenerate gas in the relativistic limit \cite{horedt2004polytropes} for polytropic EoS II.
 
In the non-relativistic limit, the ratio between pressure and density at the centre vanishes. Thus the parameter $\sigma$ gives a measure of how relevant the general relativity treatment is. Moreover, when $\sigma \rightarrow 0$ the TOV equation (\ref{ansatzcosenza}) reduces to Newtonian hydrostatic equilibrium equation \cite{Tooper1964}. Finally, a variation of anisotropic factor $C$ allows us to explore the incidence of small deviations from the isotropic condition in compact objects' stability.

\section{Results}
\label{Result}
\begin{figure}[H]
\begin{minipage}[b]{1\textwidth}
\centering
\includegraphics[width=0.75\textwidth]{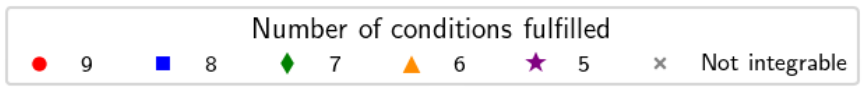}
\label{fig:my_label}
\end{minipage}\vspace{-0.5cm}
\begin{minipage}[b]{.48\textwidth}
\centering
\includegraphics[width=1\textwidth]{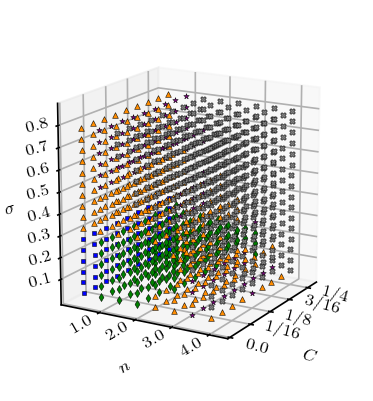}
\caption{Parametric space for the polytropic equation of state with mass density, for models with polytropic index from $n=0.5$ to $n=4.0$, and parameters $C$ and $\sigma$ varying from $0$ to $0.25$ and $0.05$ to $0.8$, respectively. None of the models fulfill all the conditions, and about half are not integrable.}
\label{EspParDenMas}
\end{minipage}
\quad
\begin{minipage}[b]{.48\textwidth}
\centering
\includegraphics[width=1\textwidth]{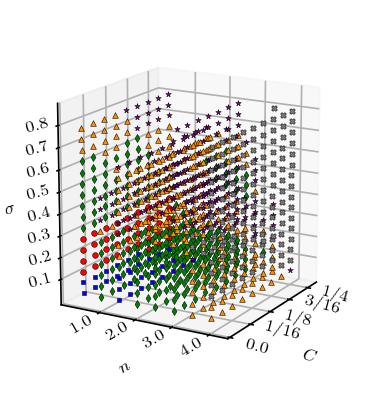}
\caption{Parametric space for the polytropic equation of state with energy density, for models with polytropic index from $n=0.5$ to $n=4.0$, and parameters $C$ and $\sigma$ varying from $0$ to $0.25$ and $0.05$ to $0.8$, respectively. Models that fulfill less than 5 conditions are not shown.}
\label{EspParDenEner}
\end{minipage}
\end{figure}

By applying acceptability conditions to these two kinds of polytropic EoS, we found that those implemented with the total energy density are more stable than when considering the baryonic mass density. As can be appreciated from figure \ref{EspParDenEner}, most models fulfil the same or a more significant number of conditions than those in figure \ref{EspParDenMas}.

It is worth mentioning that many models considering the baryonic mass density, were not able to complete the integration and it was not possible to find a physical boundary of the star such that $P(r_b)=P_{b}=0$.

\section{Conclusions}
\label{Conclusions}
We assessed the viability of relativistic hydrostatic polytropic spheres to model stable material configurations through the compliance of nine physical acceptability conditions. Two different polytropic EoS were taken into account to describe the matter within the object, differing in the type of density considered, i.e., baryonic mass density (I) or energy density (II).

Based on the parameter space obtained we conclude that models with low polytropic index $n$, i.e. highly incompressible matter, fulfill a greater number of conditions of physical acceptability. As expected, when $n \geq 3$ condition on the adiabatic index is not fulfilled \cite{horedt2004polytropes}. Also, high values of $n$ results in a breach of the adiabatic convective stability condition, namely $\rho^{\prime \prime} \geq 0$ for some value of $r$ within the object.

Regarding $\sigma$, high values for this parameter disfavors stability either breaching the strong energy condition, violating causality conditions on sound speeds or presenting cracking within the object. Thus, low values of $\sigma$ model more stable configurations. 

Small deviations from isotropic condition, i.e. $C \neq 0$, do not have greater incidence in stability. However, for some models, increasing this parameter causes fewer conditions to be fulfilled.

For many mass density models, it was not possible to finish the numerical integration, i.e. we could not meet the condition for total radii of the star such that $P(r_b)=P_{b}=0$. This result coincides with the parameter space, for bounded sources, presented in \cite{HerreraBarreto2013} for both polytropic EoS.

\vspace{0.5cm}

\ack
We are grateful for the support of the Vicerrectoría de Investigación y Extensión of the Universidad Industrial de Santander, and the financial support of COLCIENCIAS under contract No. 8863.

\section*{References}
\providecommand{\newblock}{}

\end{document}